# The Unreasonable Effectiveness of the Air-Fluorescence Technique in Determining the EAS Shower Maximum

P. Sokolsky and R. D'Avignon
Department of Physics and Astronomy
University of Utah
Salt Lake City, Utah

## Abstract

We review all existing air-fluorescence measurements of the elongation rate of extensive air showers (slope of mean EAS shower maximum (Xmax) vs log of shower energy E) above $10^{17}$ eV. We find remarkable agreement for all current and historic experiments over a 30 year period for the energy range from $10^{17}$ to $3\times10^{18}$ eV. The mean elongation rate in this energy interval is near 80 $g/cm^2$/decade. Above this energy, experiments in the Northern hemisphere are in good agreement with an average elongation rate of 48 +/- 10 $g/cm^2$/decade while Southern hemisphere experiments have a flatter elongation rate of 26 +/- 2 $g/cm^2$/decade  We point out that, given the agreement at lower energies, possible systematic reasons for this difference are unlikely. Given this, the world elongation rate data alone may indicate a composition difference of UHECR in the Northern and Southern hemisphere and thus a diversity of UHECR sources in the Northern and Southern sky.

## 1. Introduction

One of the great advances in the study of UHECR is the development of the air-fluorescence technique [1]. The ability to reconstruct the development of an extensive air shower (EAS) produced by cosmic rays at $> 10^{17}$ eV in the atmosphere has given us a much improved, essentially calorimetric, energy determination. This allows both direct measurement of the CR spectrum and the inter-calibration of the older surface array technique where only the footprint of the EAS on the surface is sampled with scintillation or water Cherenkov counters. The air-fluorescence method also gives the ability to determine, albeit with some uncertainty, the cosmic ray composition. This is done by determining the distribution of the depth of EAS shower maxima or Xmax (see [2] for an early discussion of this technique). Heavy nuclei interact early and produce showers at smaller atmospheric depths while protons interact later and have deeper Xmax. Intermediate nuclei lie between these two extremes. Thus, the fluctuations of Xmax and the actual shape of the distribution of Xmax's contain information about the composition. The limitation of this method is the necessity of comparing the observations with detailed simulations based on hadronic interaction models. These models deal with interactions well beyond current accelerator data and hence have significant uncertainties. This is reflected in the uncertainty in the inferred composition. Determining the Xmax of an EAS is also more subject to systematic uncertainties than its energy. Detailed fits to the data distributions to extract the CR composition thus suffer from multiple systematic issues.

It is possible, however, to ask a different question. Whatever the components of the cosmic ray flux may be, is this composition changing as a function of energy? Such changes may reflect propagation effects from the sources, changing acceleration efficiency at the astrophysical sources, or the appearance of different sources. A sensitive indicator of such a change in composition is the so-called elongation plot, or the dependence of the mean Xmax on log(E).  For a single component composition, it is easy to show[3] that

$$Xmax = D \ln(E/E_c)$$

Where $E_c$ is the critical energy and D depends on the particle and the hadronic model assumed.

Irrespective of the actual mixture, for a constant composition, the slope of the elongation

$$d(Xmax)/d(\log_{10}(E))$$

is constant. However, if the composition is changing over an energy interval then this slope, or elongation rate, will exhibit a corresponding change (see Fig 1 for an illustrative example). While the precise correspondence of the elongation rate to the composition is hadronic model dependent, the energy dependence of interaction parameters such as total cross section, inelasticity and multiplicity are typically logarithmic and are not expected to produce a rapid change in the elongation rate. Thus, a rapid change in the rate is most simply explained in terms of a change in composition assuming no hadronic "new physics" thresholds.

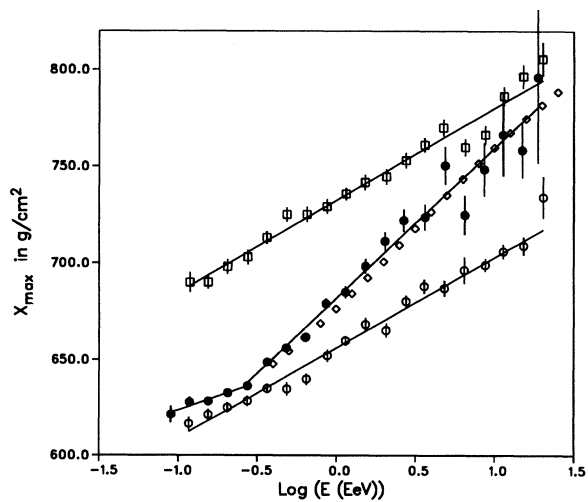

Fig. 1 Original Stereo Fly's Eye elongation rate measurement (solid circles). Top line (open squares)  is the expectation for a pure proton composition. Bottom line (open circles)  is the expectation for a pure Fe composition based on early hadronic

models. Open diamonds represent expectations for a smooth transition from Fe to protons. See [4] for details.

The major advantage of studying the UHECR elongation rate is that one can come to certain conclusions about the composition in an essentially hadronic model independent way. However, this kind of study is still sensitive to systematic effects inherent in the air-fluroescence measurement technique. It is therefore important to have multiple independent measurements to determine how well the systematic errors are controlled.

In this paper, we will compare results from all the historic air-fluorescence experiments in the Northern and Southern hemispheres, and look for consistency, or lack thereof, in reported elongation rates – in particular for evidence of a change. The eight experiments considered span a time period of 40 years and reflect a wide variety of reconstruction techniques, calibration procedures and atmospheric corrections. As we will see, somewhat incredibly, in the energy range from $10^{17}$ to $3 \times 10^{18}$ eV there is remarkable agreement about the elongation rate for Northern hemisphere experiments. Even the absolute values of average Xmax's lie well within the estimated systematic errors of 20-30 g/cm$^2$. We also present the comparison of the mean elongation rate of all the experiments in the Northern hemisphere with the result of the Auger experiment, which is the single detector in the Southern hemisphere. We will see that with the reversion to the mean of the seven Northern results, the agreement with Auger below $3 \times 10^{18}$ eV is remarkable.

## 2. The Northern Experiments
.
We briefly describe the seven experiments with elongation rate results in the North. All of these were based in the western deserts of the state of Utah, USA.

The oldest and pioneering air-fluorescence experiment was the Fly's Eye[5]. This experiment was unique in that it had full sky coverage for one of its stations. A second station with a more limited field of view was placed 3 km away to provide for stereo reconstruction. The pixel size coverage of the sky was 5 x 5 degrees, reflecting the size of the photomultipliers used. For events seen in stereo, the Xmax resolution was ~ 40 g/cm$^2$. We note that one of the advantages of stereo detection is the ability to directly measure the Xmax resolution by comparing the two independent measurements for each event. The detector ran from 1985 to 1992 and produced enough data to allow studies of UHECR spectrum and composition from $10^{17}$ eV to above $10^{19}$ eV.

The next generation experiment was the High Resolution Fly's Eye (HiRes)[6]. This detector utilized a much smaller 1 deg by 1 deg pixel size to increase the sampling of the EAS profile, increase the signal to noise ratio, and thus detect more distant events. A prototype HiRes detector with a limited field of view overlooking the CASA-MIA surface and underground muon array was first built [7]. The surface array covered a 1 km$^2$ area and was 3 km away. An underground muon array covered a similar area. This combination was the first hybrid air-fluorescence detector to attempt to reconstruct an EAS development. Events detected in coincidence by air-fluorescence and the muon array allowed accurate geometrical reconstruction, while the 1 deg by 1 deg pixel size of the HiRes prototype gave improved measurement resolution of the shower development. The resulting Xmax resolution was ~25 g/cm$^2$. Because of the limited CASA-MIA area, the effective energy range was $10^{17}$ eV to $10^{18}$ eV.

The HiRes experiment consisted of two detector stations separated by 12 km. With 1 deg by 1 deg pixel size, events seen in stereo could be reconstructed with ~ 20 g/cm$^2$ Xmax resolution from $10^{18}$ to $10^{20}$ eV. Because events could be reconstructed at distances of up to 30 km, significant attention was paid to understanding the effects of atmospheric attenuation and detailed simulation of the detector aperture.

The currently operating Telescope Array (TA) experiment[8] consists of three air-fluorescence stations with 1 deg by 1 deg pixel size, similar to HiRes, but overlooking a 700 km$^2$ surface scintillator detector array. EAS which simultaneously trigger a fluorescence detector and a minimum number of surface detectors can be reconstructed with ~ 20 g/cm$^2$ Xmax resolution for energies > $2 \times 10^{18}$ eV. For most of the data collected the FD and SD operate independently and hybrid events are found by time coincidence. Since 2010, a special hybrid trigger [9] was implemented which forces a SD readout whenever a FD is triggered. This improves the efficiency at lower energies and allows good (20 g/cm$^2$) Xmax resolution down to ~ $3 \times 10^{17}$ eV. There are two completely new design air fluorescence detector stations- Black Rock (BR) and Long Ridge (LR), while the Middle Drum (MD) incorporates original HiRes detectors. For this reason, Xmax results from the BR and LR stations are analyzed separately from the MD station, using different reconstruction programs. The hybrid trigger data is also analyzed using an independent reconstruction program.

In order to extend the air-fluorescence energy range, an additional detector called TALE was added to the MD TA fluorescence station[10]. This detector increases the acceptance for lower energy EAS (which develop higher in the atmosphere) by implementing mirrors looking at higher elevation angles, similar in concept to HiRes/MIA. In conjunction with this fluorescence detector, a special infill surface array located nearby was recently brought into operation so that hybrid reconstruction is possible for energies from $10^{16.6}$ to $10^{18.4}$ eV[11].

## 3. Reconstructing the EAS Xmax

Here we describe the methods used to find the shower maximum. Fig. 2 shows a typical shower which fell into the field of view of a TA detector. The rise and fall of the light curve is clearly seen. One can fit a functional form to the shower and extract the Xmax. Note however, that the shower is not symmetric about Xmax. A Gaussian form (which was used in the original Fly's Eye work) will give a reasonable fit around the shower maximum, but will be biased by the deep tail. A Gaiser-Hillas function incorporates this asymmetry and does a better job at reducing bias. Other functions have been tried which may have minor advantages. Nevertheless, depending on how much of the shower is seen by the detector, biases can still arise.

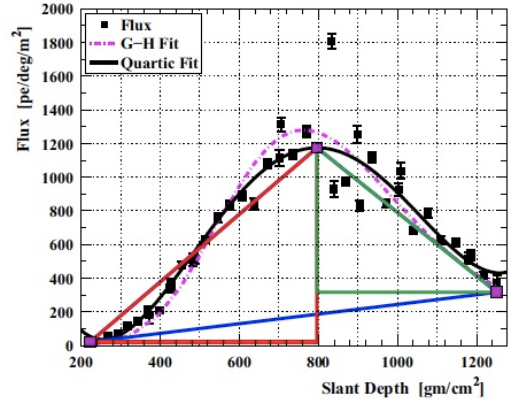

Fig. 2 A well measured TA hybrid EAS profile. Two fitting functions are presented as well as a triangular simplification of the profile used in an automated decision-tree approach to scanning for well-measured events [12].

Another issue is related to inaccurate geometrical reconstruction. Errors in geometry can produce distorted shower profiles leading to errors in Xmax determination (see Fig. 3). In some cases, shower profiles appear essentially flat, so that the fits to a function are meaningless. Similarly, the presence of aerosol layers and low clouds can distort the shower shape. These issues and how to handle them have been discussed, for instance in [12].

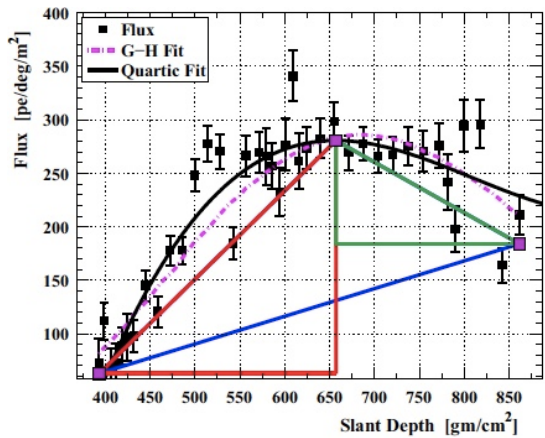

Fig. 3 Event with distorted profile due to inaccurate geometrical reconstruction [12].

There are several methods for determining how well the Xmax is found. One can simulate the detector as accurately as possible and Monte Carlo the detector response to simulated EAS. The resultant resolution is only as good as the detector simulation and may depend on the composition used in the Monte Carlo. There are important cross-checks with data however. Such a simulation also generates other observable distributions such as zenith angle, impact parameter, point of first and last observation of a shower, etc. These can be compared with real data to gain confidence in the suitability of the simulation. This was first done extensively by the HiRes collaboration ( see for example [13])

A much more direct way is to use stereo data[13]. Here the same events are seen independently by two detectors. The distribution of difference in the two values of Xmax for each event is then simply related to the detector resolution (see Fig. 4). In practice, the stereo resolution data is used to check the simulated resolution. While the tails of the distribution for real data are usually somewhat wider, the central, nearly Gaussian distributions are very similar, giving confidence in the detector simulation estimate of resolution.

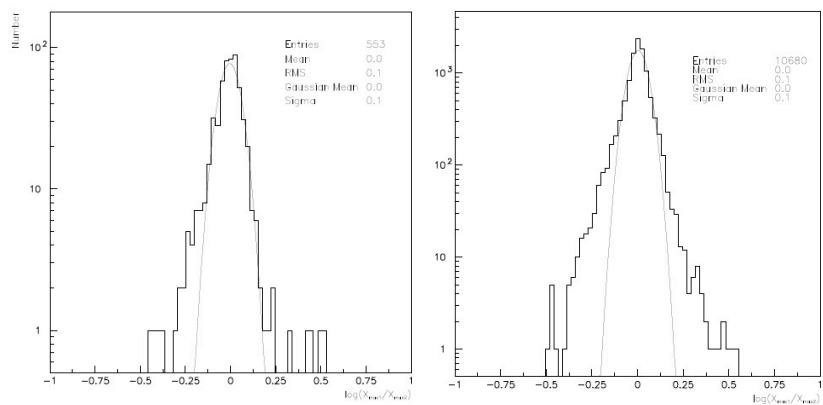

Fig. 4 Comparison of data to simulation for HiRes Stereo Xmax determination. Left: log (Xmax1/Xmax2) for simulated events. The sigma for the histogram and Gaussian fit is 0.1 Right: log (Xmax1/Xmax2) for real data. The sigma for the histogram and Gaussian fit is 0.1.

Since the Xmax resolution will depend on how well the fitting function can find a shower maximum, cuts on the data to remove poorly measured events and events at the edge of the field of view of the detector are essential. A more sophisticated approach is to filter out events that do not have an obvious Xmax in the field of view by using pattern recognition and decision tree approaches. This has been demonstrated to improve the resulting Xmax resolution significantly [12]. .

## 4. Comparison of Results for Northern Hemisphere

The seven different Northern Hemisphere experimental results, here treated as independent of each other, have different energy thresholds so that the number of mean Xmax measurements per energy bin will vary. We consider data from $10^{17}$ eV to $10^{19.2}$ eV only to guarantee that statistical errors are smaller than systematic errors and there is no significant statistical sampling bias. Fig 5 shows the individual experiment results where the statistical errors bars have been suppressed for clarity. The overall agreement as to the elongation rate is impressive. The biggest departure is the oldest result from the Fly's Eye which has deeper mean Xmax and shows a minor flattening at low energies. As discussed above, the Fly's Eye analysis used a symmetric Gaussian function to determine Xmax which can produce a energy dependent bias. Under the assumption that reversion to the mean will give the most reliable result, we form a mean and a standard deviation for each energy bin. Fig 6 shows the result where the error bars represent the standard deviation of all the experiments that contribute to a particular energy bin.

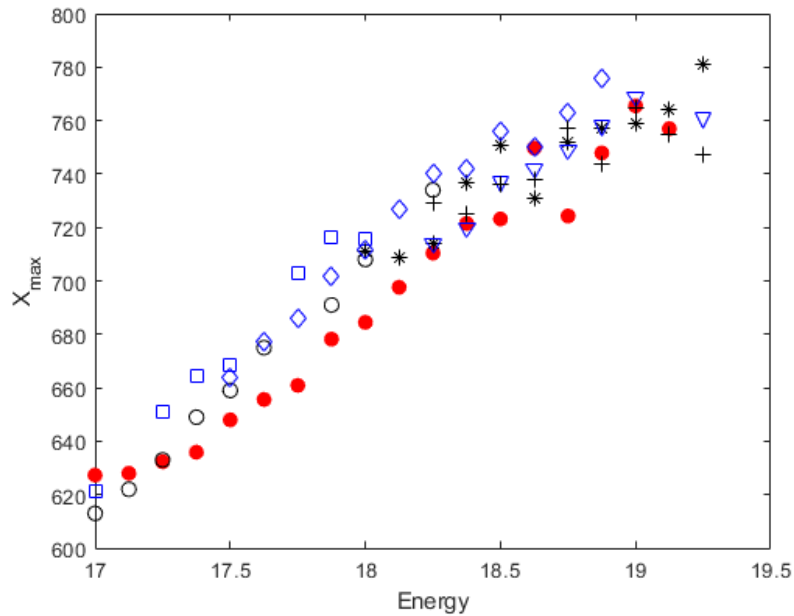

Fig. 5 Mean Xmax as function of energy for Northern hemisphere measurements. Statistical error bars have been suppressed for clarity.
Key: red circles – Fly's Eye Stereo; open circle-TALE hybrid; diamonds-TA hybrid trigger; square-HiRes-MIA; star-HiRes stereo; triangle-TA BR-LR hybrid; cross-TA-MD-hybrid;

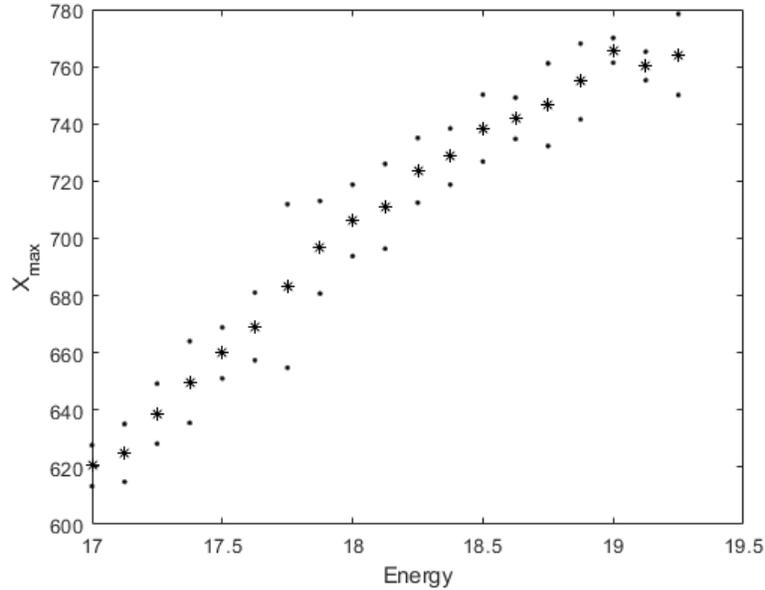

Fig. 6 Mean Xmax as function of energy averaged over all Northern data. Error bars are the standard deviation of the different experiments about the mean for each energy bin.

## 5. Results from the Southern Hemisphere

The Auger collaboration [15] has constructed a hybrid air-fluorescence surface detector array covering $\sim$ 3000 km$^2$ in the high desert of Argentina. The surface array utilizes water tanks instead of the scintillation counters used in TA. Four fluorescence detector stations overlook the ground array. The fluorescence detectors have a similar pixel size to TA and HiRes ($\sim$ 1 deg by 1 deg). The collaboration has now published high statistics results on Xmax distributions[16]. The original results were limited to the higher energies, but recent detector extensions to lower energies, using similar strategies to what was used for TA and TALE, allow the collaboration to produce a unified elongation rate from $2 \times 10^{17}$ to the highest energies. The result is shown in Fig. 12. There is a clear break in the elongation rate at $\sim 3 \times 10^{18}$, which is interpreted as indicating the beginning of a departure from a light, mainly protonic composition to a mixed composition including He and CNO. This is corroborated by detailed fits to the Xmax distributions themselves. The Xmax resolution of 15 – 20 g/cm$^2$ is similar to what is achieved in the North by TA.

## 6. Comparison of North and South Elongation Rates.

Fig 7 shows the North and South elongation rates. There is strikingly good agreement in slopes from $10^{17}$ to $\sim 3 \times 10^{18}$ eV. There is an $\sim$ 25 g/cm$^2$ systematic shift between the measurements, consistent with the estimated systematic errors of $\sim$20 g/cm$^2$ for each experiment. Fig 8 shows the same data with the Auger results shifted down by 25g/cm$^2$. The lower energies show a remarkable agreement with elongation rate of ~85 g/cm²/decade for the North and 79.1 g/cm²/decade for Auger. The slopes above 3x10¹⁸ eV are different however. Auger determines a rate of 26 +/- 2 while the average of the Northern experiments gives a rate of 47.8 g/cm²/decade with a standard deviation of 10.4.

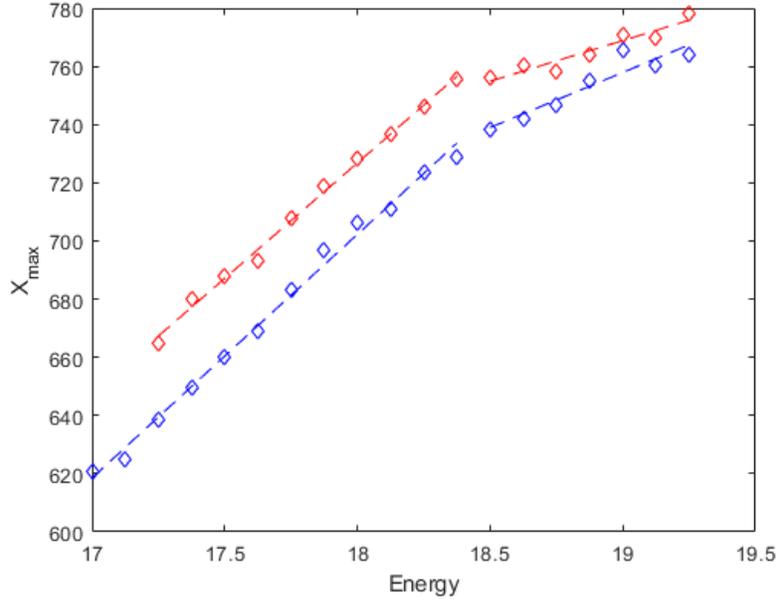

Fig. 7 North and South Elongation Rates. Red points are Auger data. Blue points are the mean Northern data. Dashed lines are linear fits with an assumed break at 3 x 10¹⁸ eV. Error bars have been suppressed for clarity but are smaller than the ~20g/cm² systematic errors. Auger data points have been slightly interpolated in energy to correspond to the binning used here.

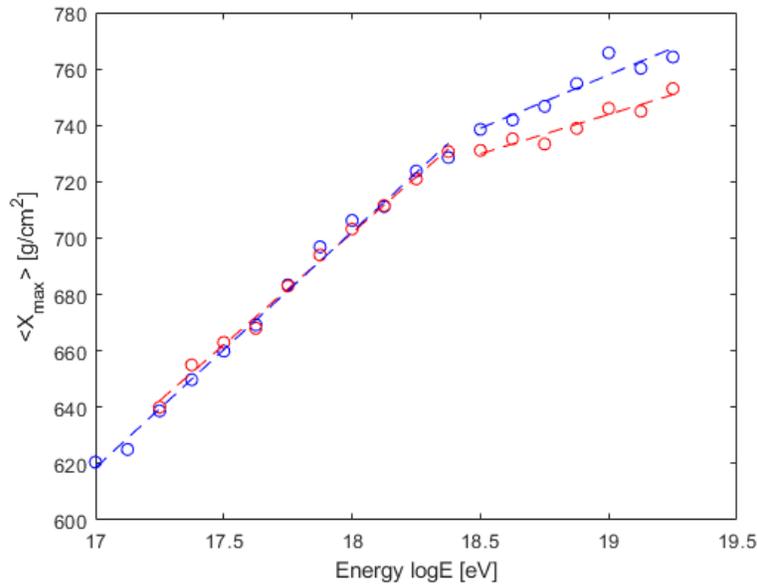

Fig. 8 North and South elongation rates after 25 g/cm² shift.

## 7. Discussion

There has been a divergence of interpretation of Xmax data at the highest energies between Auger and the various Northern experiments for some time. Detailed studies of Xmax distribution data shapes in the North, particularly the most recent TA hybrid results [14] have been most easily understood as being protonic, or nearly so, with some admixture of He and CNO possible at energies above $5 \times 10^{18}$ eV. Auger data above $3 \times 10^{18}$ when analyzed in a similar way tend to prefer a heavier composition. Until recently, comparisons, done by the Joint Composition Working Group[17] have only been made in a limited energy region of $2 \times 10^{18}$ to $2 \times 10^{19}$ eV. Reasonable agreement in the Xmax distributions themselves (if not the hadronic model dependent interpretations) could be found if one shifted either experiment's results bin by energy bin by between 5 and 20 $g/cm^2$. However, the elongation rates between the two experiments above $2 \times 10^{18}$ eV remained different, though the evidence for this was weakened by the small energy range available and lack of statistics at the highest energies.

The present analysis using all available data paints a clearer picture. At lower energies, the agreement between North and South is highly reassuring. But the much increased lever arm available shows a real difference in measured elongation rate, beginning at $3 \times 10^{18}$ eV. If we assume there are no unexplored systematic effects, then this would indicate that the composition of cosmic rays in the Northern and Southern hemisphere begins to diverge at this energy, remaining relatively light in the North and getting heavier more rapidly in the South. The sources of the UHECR in the North and South could be different.

### 7.1 Systematics

Before coming to such an important conclusion, we must exclude possible systematic biases between North and South. Note that because we are talking about elongation rates, overall biases do not matter – only energy dependent ones are important. Because of the excellent agreement between North and South below $3 \times 10^{18}$ eV, the growth of biases in event geometry reconstruction, correction for atmospheric effects, and Xmax determination with energy, must be nearly identical, if they are there at all. Similarly, the change of detector aperture with respect to zenith angle must be very similar as a function of energy in this energy range.

The Northern TA hybrid, TA stereo and HiRes stereo data begins right where the break in the elongation rate occurs and the divergence with Auger begins. Is it then possible that these experiments have a different bias than the lower energy Northern experiments and the elongation rate break is an experimental artifact? Fortunately, this can be ruled out by the new TA hybrid trigger and TALE hybrid data [9] [11] (Fig. 9) which spans from $10^{17}$ to $10^{18.8}$, covering the break point and in excellent agreement at higher and lower energies with the other more limited energy range Northern data.

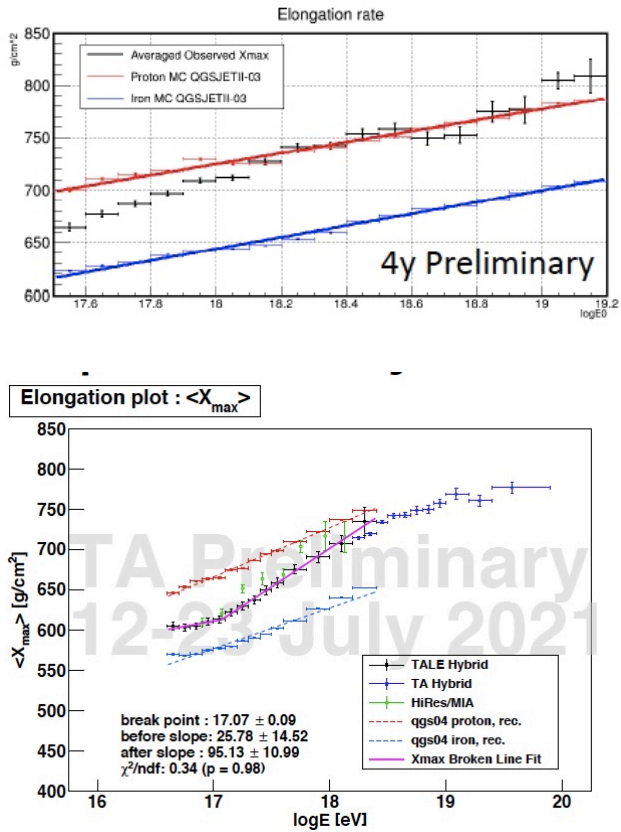

Figure 9. Top: TA-hybrid trigger elongation rate. Bottom: TALE-hybrid elongation rate[9][11].

What remaining systematic could possibly account for the slope difference above $3 \times 10^{18}$ eV? The experiments in the North and South employ rather different sets of cuts in the data. The Auger analysis uses hard cuts on EAS variables to insure that the data lie in a region of flat detector acceptance. Corrections for the missing tail events are then applied. The data should then have no detector bias and can be directly compared to unbiased simulations. The TA and HiRes analysis only applies cuts sufficient to guarantee good Xmax resolution. Showers with deep Xmax are not removed. The result is an Xmax dependent variation in detector efficiency. This is handled in the comparison to Monte Carlo simulations by requiring that simulated data be treated exactly like the actual data. When this is done, the data and simulated data elongation rates match quite well (Fig. 10 shows the comparison for TA-BR-LR hybrid data). In addition, one can check how much the detector-independent (thrown) simulated elongation rate and the reconstructed simulated elongation rate differ. While there are overall shifts of mean Xmax position of up to ~ 10 g/cm², there is no evidence for a significant energy dependence in the shift [14](Fig. 11).

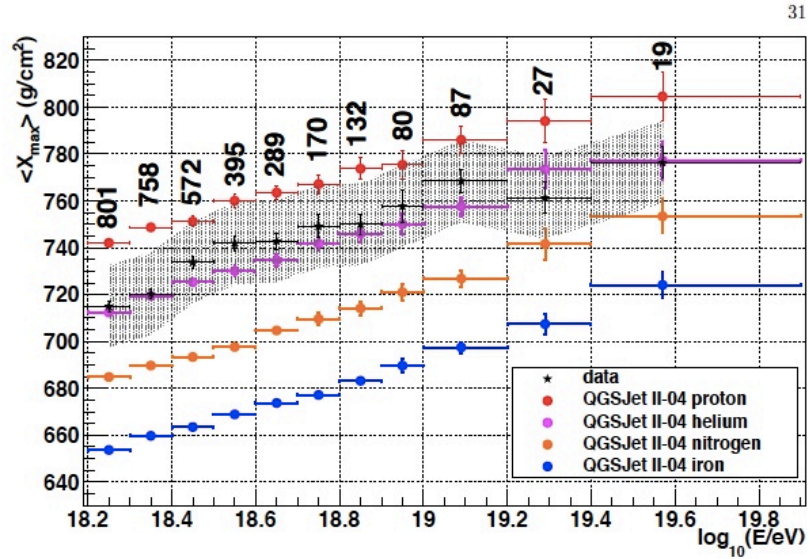

Fig. 10 TA-hybrid elongation rate compared to simulated predictions for four different representative elements. The simulated events undergo identical selection cuts as the real data[14].

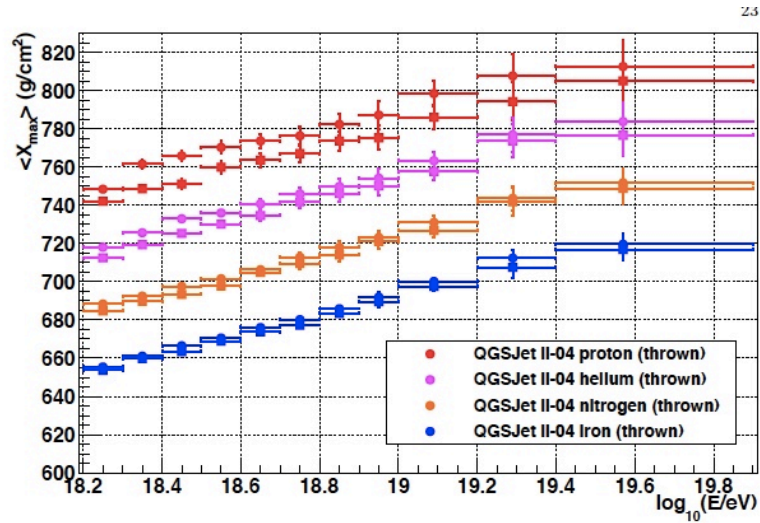

Fig 11. TA-hybrid simulated elongation rates for individual elements. Circles correspond to the detector independent (thrown) events while squares give the result after detector triggering and event reconstruction simulation[14].

Events reconstructed in stereo have better acceptance at large zenith angles (corresponding to deeper Xmax) than hybrid events, yet there is no signIficant difference in elongation rates between HiREs stereo, TA stereo and TA hybrid data so that variations in zenith angle acceptance at the level seen is not significant.

On the other hand, the Auger elongation rate (Fig. 12), resulting from cuts that remove detector bias, when compared with simulations that do not include detector effects show a very different picture, with clear indication of a movement away from a constant proton dominated composition.

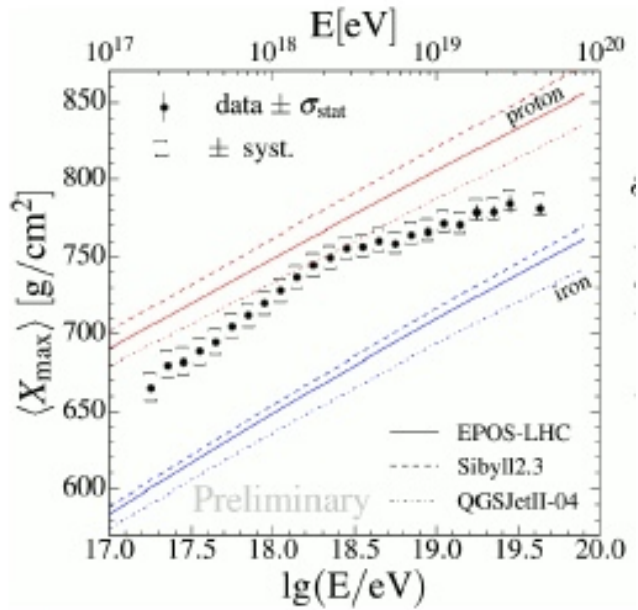

Fig, 12 Auger elongation rate compared to experimentally unbiased predictions of proton and Fe elongation rates[16].

## 8. Conclusions

The remarkable agreement in elongation rate for eight different measurements gives strong impetus to using the divergence above $3 \times 10^{18}$ eV as a tool for exploring differences in North/South composition. However, a consensus needs to be achieved that there are no improperly understood systematic effects in this region. In this regard, [18] gives a historical account of North/South measurements from a different perspective, but comes to a similar conclusion about the existence of a break near $3 \times 10^{18}$ eV.

Since the important difference is in the slope, these systematics, if they exist, must increase with energy, but only above $3 \times 10^{18}$ eV. They must either increasingly push Xmax to smaller values (the Auger case), or increasingly push Xmax to larger values (the Northern case). Two candidates for such effects are: cuts which increasingly throw out deep Xmax, and increasingly poor Xmax resolution resulting in an increasing deep Xmax tail not modeled in the simulations. Neither of these possibilities is supported by the work of the experimental groups. Nevertheless, the Joint TA-Auger Composition Working Group should press on with further elucidation of such possible effects. Absent such effects, the divergence in composition North and South joins the emergence of different anisotropies (Cen A and Starburst Galaxies in the South [19]; the Hot Spot and the Perseus-Pices supercluster enhancement in the North [20]) as a strong indication of the diversity of cosmic ray sources at the highest energies.

It is remarkable that a technique that is based on the observation of the emission of ~4 photons/particle/m at distances of up to 30 km, in the presence of significant sky noise, and which has to take into account molecular and aerosol scattering in the atmosphere as well as the stability and calibration of thousands of pixels can produce such reliable and reproducible experimental results. The pioneers of this idea ~60 years ago (Chudakov [21] in the USSR and others in Japan and the United States) would be pleased at what their original insight has brought forth.